\documentclass{llncs}
\usepackage[normalem]{ulem}

\usepackage{noitemsep}
\usepackage{latexsym,
epsfig,
graphicx,
xspace,
cite,
amssymb,
amsmath,
amsfonts}
\usepackage{wrapfig}

\makeatletter
\newif\if@restonecol
\makeatother

\usepackage[ruled,figure,linesnumbered,noresetcount]{algorithm2e}

\newcommand{\comment}[1]{}  
\newcommand{\old}[1]{} 

\newcommand{\affa}{$^{1}$}
\newcommand{\affb}{$^{2}$}
\newcommand{\affc}{$^{3}$}
\newcommand{\affd}{$^{4}$}

\newcommand{\ignore}[1]{}  
\newcommand{\aparagraph}[1]{{\noindent \bf \emph{#1}}}

\newcommand{\squishlist}{
  \begin{list}{$\bullet$}
   {
     \setlength{\itemsep}{0pt}
     \setlength{\parsep}{0pt}
     \setlength{\topsep}{0pt}
     \setlength{\partopsep}{0pt}
     \setlength{\leftmargin}{1.5em}
     \setlength{\labelwidth}{1em}
     \setlength{\labelsep}{0.5em} } }
\newcommand{\squishend}{
   \end{list}  }

\newcommand{\scicache}{{\sf Delta}\xspace}
\newcommand{\river}{\scicache}

\newcommand{\benefit}{{\small \sf Benefit}\xspace}
\newcommand{\vcover}{{\small \sf VCover}\xspace}
\newcommand{\replica}{{\small \sf Replica}\xspace}
\newcommand{\nocache}{{\small \sf NoCache}\xspace}
\newcommand{\soptimal}{{\small \sf SOptimal}\xspace}
\newcommand{\stale}{{\small \sf stale}\xspace}
\newcommand{\fresh}{{\small \sf fresh}\xspace}

\newcommand{\lazy}{{\small \sf lazy}\xspace}
\newcommand{\loadmanager}{\ensuremath{\small \mathsf{Load\-Manager}}\xspace}
\newcommand{\updatemanager}{\ensuremath{\small \mathsf{Update}\-\mathsf{Manager}}\xspace}

\newcommand{\aobj}{\ensuremath{\mathcal{A}_{\rm obj}}\xspace}
\newcommand{\Aobj}{\ensuremath{\mathcal{A}_{\rm obj}}\xspace}

\newenvironment{proofsketch}{\emph{Proof Sketch.}\
}{\qed\par\vskip 4mm\par}

\title{A Dynamic Data Middleware Cache for Rapidly-growing Scientific Repositories}
\author{
{Tanu Malik\affa, Xiaodan Wang\affb, Philip Little\affc, \\ 
Amitabh Chaudhary\affc,  Ani Thakar\affd}
\institute{
\affa\hspace{0.12cm}Cyber Center, Purdue University\\
tmalik@cs.purdue.edu\\
\affb\hspace{0.12cm}Dept. of Computer Science, \affd\hspace{0.12cm}Dept. of Physics and Astronomy, \\
Johns Hopkins University \\
xwang@cs.jhu.edu, thakar@pha.jhu.edu \\
\affc\hspace{0.12cm}Dept. of Computer Science and Engg., \\
University of Notre Dame\\
achaudha, plittle@cse.nd.edu  \\
}}

\begin{document}
\maketitle

\begin{abstract}

Modern scientific repositories are growing rapidly in size. 
Scientists are increasingly interested in viewing the latest data as part of query results. 
Current scientific middleware cache systems, however, assume repositories are static. Thus, they cannot answer scientific queries with the latest data. The queries, instead, are routed to the repository until data at the cache is refreshed. 
In data-intensive scientific disciplines, such as astronomy, indiscriminate query routing or data refreshing often results in runaway network costs. 
This severely affects the performance and scalability of the repositories and makes poor use of the cache system.
We present \river a dynamic data middleware cache system for rapidly-growing scientific repositories. 
\river's key component is a decision framework that adaptively \emph{decouples} data objects---choosing to keep some data object at the cache, when they are heavily queried, and keeping some data objects at the repository, when they are heavily updated. 
Our algorithm profiles incoming workload to search for optimal data decoupling that reduces network costs. It leverages formal concepts from the network flow problem, and is robust to evolving scientific workloads.  
We evaluate the efficacy of \river, through a prototype implementation, by running query traces collected from a real astronomy survey. 
\end{abstract}

\section{Introduction}
\label{sec:Intro}
\vspace{-5pt}


  Data collection in science repositories is undergoing a transformation. This is remarkably seen in astronomy. Earlier surveys, such as the Sloan Digital Sky Survey (SDSS) \cite{sdss, SGT+02} collected data at an average rate of 5GB/day. The collected data was added to a database repository through an off-line process; the new repository was periodically released to users. However, recent surveys such as the Panoramic Survey Telescope \& Rapid Response System (Pan-STARRS) \cite{panstarrs} and the Large Synoptic Survey Telescope (LSST) \cite{lsst_web} will add new data at an average rate considerably more than 100 GB/day! Consequently, data collection pipelines have been revised to facilitate continuous addition of data to the repository 
  \cite{kaiser2002panstarrs}. Such a transformation in data collection impacts how data is made available to users when remote data middleware systems are employed.
  
  Organizations deploy data middleware systems  to improve data availability by reducing access times and network traffic \cite{shoshani2004storage,dbcache}. A critical component of middlewares systems are database caches that store subsets of repository data in close proximity to users and answer most user queries on behalf of the remote repositories. Such caches worked well with the old, batch method of data collection and release. 
But when data is continuously added to the repositories, cached copies of the data rapidly become stale. Serving stale data is unacceptable in sciences such as astronomy where users are increasingly interested in the latest observations.
Latest observations of existing and new astronomical bodies play a fundamental role in time-domain studies and light-curve analysis \cite{kaiser:11:timedomain,protopapas2005-1lightcurve}.
To keep the cached copies of the data fresh, the repository could continuously propagate updates to the cache. 
 But this results in runaway network costs in data-intensive applications.


Indeed, transforming data middlewares to cache dynamic subsets of data for rapidly growing scientific repositories is a challenge. Scientific applications have dominant characteristics that render dynamic data caches proposed for other applications untenable. Firstly, scientific applications are data intensive. In PAN-STARRS for instance, astronomers expect daily data additions of atleast 100GB and a query traffic of 10TB each day. Consequently a primary concern is minimizing network traffic. Previously proposed dynamic data caches for commercial applications such as retail on the Web and stock market data dissemination have a primary goal of minimizing response time and minimizing network traffic is orthogonal. Such caches incorporate latest changes by either (a) invalidating subsets of cached data and then propagating updates or shipping queries, or (b) by proactively propagating updates at a fixed rate. Such mechanisms are blind to actual queries received and thus generate unacceptable amounts of network traffic.

Secondly, scientific query workloads exhibit a constant evolution in the queried data objects and the query specification; a phenomenon characteristic of the serendipitous nature of science \cite{singh:traffic,wang2007workload,malik_black-box}. The evolution often results in entirely different sets of data objects being queried in a short time period. In addition, there is no single query template that dominates the workload. 
Thus, it is often hard to extract a representative query workload. For an evolving workload, the challenge is in making robust decisions---that save network costs and remain profitable over a long workload sequence. Previously proposed dynamic data caches often assume a representative workload of point or range queries \cite{deolasee,candan,dar_semantic}. 

In this paper, we present \river a dynamic data middleware cache system for rapidly growing scientific repositories. \river addresses these challenges by incorporating two crucial design choices: 

(A) \label{designA} Unless a query demands, no new data addition to the repository is propagated to the cache system, If a query demands the latest change, \river first invalidates the currently available stale data at the cache. The invalidation, unlike previous systems \cite{shoshani2004storage,chandan01cacheinvalidation}, is not followed by indiscriminate shipping of queries or updates; \river incorporates a decision framework that continually compares the cost of  propagating new data additions to the cache with the cost of shipping the query to the server, and adaptively decides whether it is profitable to ship queries or to ship updates.  

(B) \label{designB} In \river, decisions are not made based on assuming some degree of workload stability.  Often frameworks assume prior workload to be an indicator of future accesses. Such assumptions of making statistical correlation on workload patterns lead to inefficient decisions, especially in the case of scientific workloads that exhibit constant evolution. 

To effectively implement the design choices, the decision framework in \river decouples data objects; it chooses to host data objects for which it is cheaper to propagate updates, and not host data objects for which it is cheaper to ship queries. The decoupling approach naturally minimizes network traffic. If each query and update accesses a single object, the decoupling problem requires simple computation: if the cost of querying an object from the server exceeds the cost of keeping it updated at the cache, then cache it at the middleware, otherwise not. However, scientific workloads consist of SQL queries that reference multiple data objects. A general decoupling problem consists of updates and queries on multiple data objects. We show how the general decoupling problem is a combinatorial optimization problem that is NP-hard.

We develop a novel algorithm, \vcover, for solving the general decoupling problem. \vcover is an incremental algorithm developed over an offline algorithm for the network flow problem. \vcover minimizes network traffic by profiling costs of incoming workload; it makes the best network cost optimal decisions as are available in hindsight. It is robust to changes in workload patterns as its decision making is grounded in online analysis. The algorithm also adapts well to a space constrained cache. It tracks an object's usage and makes load and eviction decisions such that at any given time the set of objects in cache satisfy the maximum number of queries from the cache.
We demonstrate the advantage of \vcover for the data decoupling framework in \river by presenting \benefit, a heuristics based 
greedy algorithm that can also decouple data objects. Algorithms similar to \benefit  
are commonly employed in commercial dynamic data caches \cite{labrinidis-dt_frshnss_wb_srvrs04}.


We perform a detailed experimental analysis to test the validity of the decoupling framework in real astronomy surveys. 
We have implemented both \vcover and \benefit and experimentally evaluated their performance using more than 3 Terabyte of 
astronomy workload collected from the Sloan Digital Sky Survey. We also compare them against three yardsticks: \nocache, \replica and \soptimal. Our experimental results show that Delta (using \vcover) reduces the traffic by
nearly half even with a cache that is one-fifth the size of the server repository. Further, \vcover
outperforms \benefit by a factor that varies between 2-5 under different conditions.
It's adaptability helps it maintain a steady performance in the scientific real-world, where
queries do not follow any clear patterns.

\noindent{\bf RoadMap:} We discuss related works in Section \ref{sec:related}. The science application, \emph{i.e.,} the definition of  data objects, and the specification of queries and updates is described in Section \ref{sec:problemformulation}. In this Section, we also describe the problem of decoupling data objects between the repository and the cache. An offline approach to the data decoupling problem is described in Section \ref{sec:offline}. In Section \ref{sec:vcover}, we describe an online approach, \vcover, to the data decoupling problem. This approach does not make any assumptions about workload stability. An alternative approach, \benefit, which is based on workload stability is described in Section \ref{sec:benefit}. We demonstrate the effectiveness of \vcover over \benefit in Section \ref{sec:exp}. Finally, in Section \ref{sec:conclusion} we present our conclusions and future work.

\section{Related Work}
\label{sec:related}
\label{sec:relatedwork}

The decoupling framework proposed in \river to minimize network traffic and
improve access latency is similar to the 
hybrid shipping model of Mariposa \cite{Ston94, Ston96}.  In Mariposa,
processing sites either ship data to the client or ship query to the server for processing. This
paradigm has also been explored in OODBMSs \cite{LRV, Deux} 
More recently, applications use content distribution networks for efficiently delivering data over the Internet \cite{peng2004cdn}. However, none of these systems consider propagating updates on the data \cite{olston-cdn}. In \river the 
decoupling framework includes all aspects of data mobility, which
include query shipping, update propagation and data object loading.

Deolasee et al.\ \cite{deolasee} consider a
proxy cache for stock market data in which, adaptively, either updates to a data object are pushed by the server
or are pulled by the client. Their method is limited primarily to single valued objects, such as stock prices and point queries. 
The tradeoff between query shipping and update propagation are explored in
online view materialization systems \cite{LR00,labrinidis-dt_frshnss_wb_srvrs04}. 
The primary focus is on minimizing response time to satisfy currency of queries.
In most systems an unlimited cache size is assumed. 
To compare, in \benefit we have developed an algorithm that uses workload heuristics similar to the algorithm developed in \cite{labrinidis-dt_frshnss_wb_srvrs04}. The algorithm minimizes network traffic instead of response time.
Experiments show that such algorithms perform poorly on scientific workloads.

More recent work \cite{huang1994dcc} has focused on minimizing the network traffic. However the problem is focused on 
communicating just the current value of a single object. As a result the proposed algorithms do not scale for scientific repositories in which objects have multiple values.
Alternatively, Olston et al. \cite{Olston01precision,Olston02-1} consider the precision-based approach to reduce network costs. In their approach,
users specify precision requirements for each query, instead of
currency requirements or tolerance for staleness. In scientific
applications such as the SDSS, users have zero tolerance for
approximate or incorrect values for real attributes as an imprecise
result directly impacts scientific accuracy. 

Finally, there are several dynamic data caches that maintain currency of cached data. 
These include DBProxy \cite{amiri_dbproxy:_2003}, DBCache \cite{dbcache}, MTCache
\cite{guo_support_2004} and TimesTen \cite{Time00} and the more recent
\cite{garrod_scalable}. These systems provide a comprehensive dynamic data caching
infrastructure that includes mechanisms for shipping queries and
updates, defining data object granularity and specifying currency
constraints. However, 
they lack a decision framework that adaptively chooses between 
query shipping and update propagation and data loading; 
any of the data communication method is deemed sufficient for query currency.

\section{\river: A Dynamic Data Middleware Cache}
\label{sec:problemformulation}
\begin{figure*}[ht]
  \centering
  \includegraphics[width=0.9\textwidth]{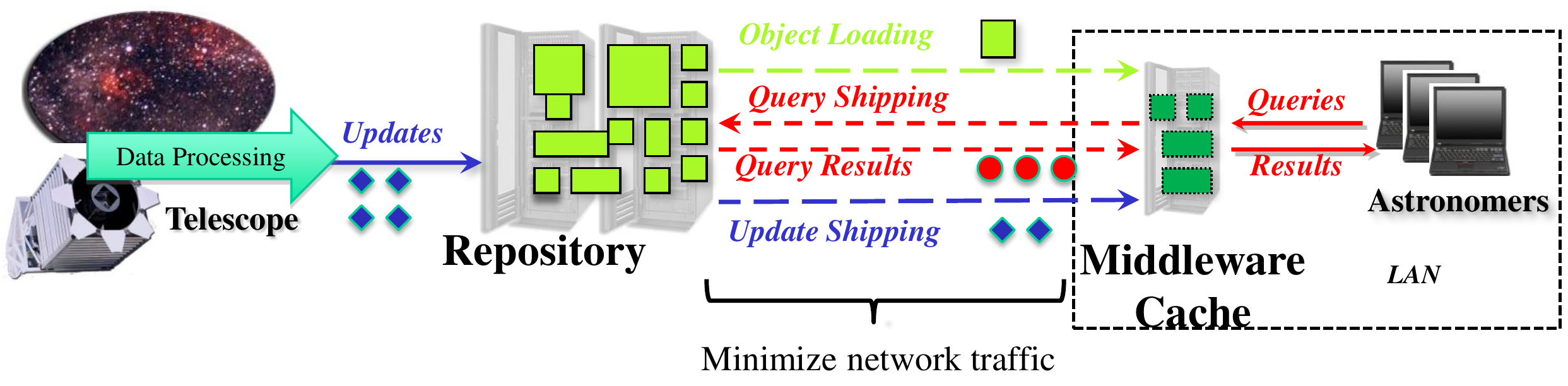}
  \caption{The {\sc \scicache} architecture.}
  \label{fig:architecture}
\end{figure*}
We describe the architectural components in \river and how data is exchanged between the server repository and the cache (See Figure~\ref{fig:architecture}). Data of a scientific repository is stored in a relational database system. While the database provides a natural partition of the data in the form of tables and columns, often spatial indices are available that further partition the data into ``objects'' of different sizes. 
A rapidly-growing repository receives updates on these data objects from a data pipeline . The updates, predominantly, insert data into, and in some cases, modify existing data objects. Data is never deleted due to archival reasons. We model a repository as a set of data objects $S = {o_1,\ldots,o_N}$. Each incoming update $u$ affects just one object $o(u)$, as is common in scientific repositories. 

Data objects are stored at the cache of the middleware system to improve data availability and reduce network traffic. The cache is located along with or close to the clients, and thus ``far'' from the repository. The cache has often much less capacity than the original server repository and is thus space-constrained. We model the cache again as a subset of data objects $C = {o_1,\ldots,o_n}$. The objects are cached in entirety or no part of it is cached. This simplifies loading 
of objects. Objects at the cache are invalidated when updates arrive for them at the server. Each user query, $q$, is a read-only SQL-like query that accesses data from a set of data objects $B(q)$. The cache answers some queries on the repository's behalf. Queries that cannot be answered by the cache are routed to the repository and answered directly from there. To quantify its need for latest data, queries may include user or system specified currency requirements in form of a \emph{tolerance for staleness}
$t(q)$, defined as follows: Given $t(q)$, an answer to $q$ must incorporate all updates
received on each object in $B(q)$ except those that arrived within the
last $t(q)$ time units. This is similar to the syntax for specifying $t(q)$ as described in \cite{guo_support_2004}.
The lower the tolerance, the stricter is the user's need for current data.

To satisfy queries as per their tolerance for staleness, the cache chooses between the three available communication mechanisms: 
(a) \emph{update shipping}, (b) \emph{query shipping}, and (c) \emph{object loading}. To ship updates, the system sends an update specification including
any data for insertion and modification to be applied on the cached data objects. In shipping queries, the system redirects the query to the repository. The up-to-date result is then sent directly to the client. Through the object loading mechanism, the cache loads objects not previously in the cache, provided there is available space to load. The three mechanisms differ from each other
in terms of semantics and the cost of using them. For instance, in update shipping only the newly inserted tuples to an object are shipped whereas in object loading the entire data object (including the updates) is shipped. 

The system records the the cost of using each of the data communication mechanism. In \river, we have currently focused on the 
network traffic costs due to the use of these mechanisms. Latency costs are discussed in Section \ref{sec:loadmanager}.
Network traffic costs are assumed proportional to the size of the data being communicated. Thus
when an object $o$ is loaded, a load cost, $\nu(o)$, proportional to the object's
size is added to the total network traffic cost. For each query $q$ or update $u$ shipped a
network cost $\nu(q)$ or $\nu(u)$ proportional to the size of $q$'s result or the size of data content in $u$
is added respectively. 
The proportional assumption relies on networks exhibiting linear cost scaling with object
size, which is true for TCP networks when the transfer size
is substantially larger than the frame size \cite{stevens}. The quantitative difference in costs between the three mechanisms is described through an example in Section \ref{sec:offline}.

In \river the difference, in terms of cost, of using each communication mechanism leads to the formulation of the data decoupling problem. We first describe the problem in words and then for ease of presentation also provide a graph-based formulation.

\vspace{10pt}
\aparagraph{The data decoupling problem:} In the data decoupling problem we are given the set of objects on the repository, the online sequence of user queries at the cache, and the online sequence of updates at the repository. The problem is to decide which objects to load into the cache from the repository, which objects to evict from the cache, which queries to ship to the repository from the cache, and which updates to ship from the repository to the cache such that (a) the objects in cache never exceed the cache size, (b) each query is answered as per its currency requirement, and (c) the total costs, described next, are minimized.  The total costs are the sum of the load costs for each object loaded, the shipping costs for each query shipped, and the shipping costs for each update shipped.

\aparagraph{Graph-based formulation:}  The decoupling problem given a set of objects in the cache, can be visualized through a graph 
$G(V,E)$ in which the vertices of the graph are queries, updates and data objects. 
Data object vertices are of two kinds: one that are in cache, and ones that are not in cache. In the graph, an edge is drawn between
(a) a query vertex and an update vertex, if the update affects the query, and 
(b) a query vertex and a data object vertex, if the object is not in cache and is accessed by the query. 
In the graph, for presentation, edges between an update vertex and an object not in cache and query vertex and an object in cache are not drawn. 
Such a graph construction helps to capture the relationships between queries, updates and objects. 
We term such a graph as an \emph{interaction} graph as the edges involve a mutual decision as to which 
data communication mechanism be used. The decoupling problem then corresponds to determining which data communication 
to use since the right combination minimizes network traffic costs.

The interaction graph is fundamental to the data decoupling problem as algorithms can
determine an optimal decoupling or the right combination of data communication mechanisms to use. 
The algorithms must also be robust to 
incoming query and update workloads since slight changes in the workload lead to entirely different data communication mechanisms becoming optimal (See design choices \ref{designA} and \ref{designB} in Section \ref{sec:Intro}).
This is demonstrated in the next subsection through an example. The example demonstrates that different choices become optimal
when the entire sequence of queries, updates and their accompanying costs are known in advance, and there are 
no cache size constraints.



\subsection{Determining off-line, optimal choices} 
\label{sec:offline}

The objective of an algorithm for a data decoupling problem is to make decisions about which queries to ship, which updates to ship, and which objects to load such that the cost of incurred network traffic is minimized. This is based on an incoming workload pattern. The difficulty in determining the optimal combination of decision choices arises because a slight variation in the workload, \emph{i.e.,} change in cost of shipping queries or shipping updates or the query currency threshold, results in an entirely different combination being optimal. This is best illustrated through an example. 

\begin{figure}[h]
  \centering
  \includegraphics[height=2.65in]{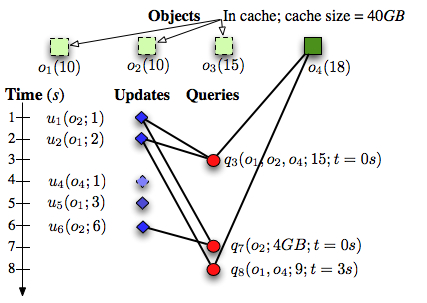}
  \caption{An decoupling graph for a sample sequence. The notation:
    $o_1(10)$ implies size of object $o_1$ is 10 GB, which is also
    its network traffic load cost. $u_1(o_2,1)$ implies update $u_1$
    is for object $o_2$ and has a network traffic shipping cost of
    $1GB$.  $q_3(o_1,o_2,o_4;15;t=0s)$ implies query
    $q_3$ accesses objects $o_1$, $o_2$, and $o_4$; has a network
    traffic shipping cost of $15GB$; and has no tolerance for staleness. }
  \label{fig:ig-example}
\end{figure}

In the Figure\ref{fig:ig-example}, among the four data objects $o_1, o_2, o_3, o_4$, objects $o_1$, $o_2$, $o_3$ have been loaded into the cache (broken lines) and object $o_4$ is currently not in cache (solid lines), 
and will be available for loading if there is space.
Consider a sequence of updates and queries over the next eight seconds. Based on the graph formulation of the decoupling problem, we draw edges between queries, updates and the objects. Query $q_3$ accesses the set {$o_1$, $o_2$, $o_4$} and has edges with
$o_4$, $u_1$, and $u_2$. since either $q_3$ is shipped or 
the use of other data communication mechanism is necessary, namely $o_4$ is loaded, and $u_1$ and $u_2$ are
shipped. For $q_7$, since object $o_2$ is in cache, the dependence is only on updates $u_1$ and $u_6$ both of which are necessary to
satisfy $q_7$'s currency constraints. $q_8$ accesses set {$o_1$, $o_4$} and so has edges with $o_4$ and $u_2$. Note, it does not have edges with $u_4$ because shipping $u_4$ is not necessary till object $o_4$ is loaded into the cache. 
If $o_4$ was loaded before $u_4$ arrives, then an edge between $u_4$ and $q_8$ will be added.  
There is no edge with $u_5$ since the tolerance for staleness of $q_8$ permits an answer from the cached
copy of $o_1$ that need not be updated with $u_5$. 

In the above example, the right choice of actions is to evict $o_3$ and load $o_4$ at the very beginning. Then at appropriate steps ship updates $u_1$, $u_2$, and $u_4$, and the query $q_7$.  This will satisfy all currency requirements, and incur a network traffic cost of $26 GB$.  Crucial to this being the best
decision is that $q_8$'s tolerance for staleness allows us to omit shipping
$u_5$.  If that were not the case, then an entirely different set of choices become optimal which include
not loading $o_4$, and just shipping queries $q_3$,
$q_7$, and $q_8$ for a network cost of $28 GB$. Such a combination minimizes network traffic.

In the above example, if we focus only on the internal interaction graph of cached objects, which is formed with nodes $u_1$, $u_6$ and $q_7$, we observe that determining an optimal decision choice corresponds to finding the minimum-weight vertex cover \cite{corman1990ia,wolfram-vc} on this subgraph. For lack of space, we omit a description on minimum-weight vertex cover and directly state the correspondence through the following theorem:

\begin{theorem}[Min Weight Vertex Cover]
Let the entire incoming sequence of queries and updates in the
internal interaction graph $G$ be known in advance.  Let $VC$ be the
minimum-weight vertex cover for $G$.  The optimal choice is to ship
the queries and the updates whose corresponding nodes are in $VC$.
\end{theorem}
\begin{proofsketch} If the entire incoming sequence of queries and
  updates is known we claim the optimal choice is: (1) choose such
  that every query $q$ is either shipped or all updates interacting
  with $q$ are shipped, and (2) make a choice such that the total
  weights of nodes chosen for shipping is minimized.  This is true
  since the sum of the weights of nodes chosen is equal to the actual
  network traffic cost incurred for the choice. This corresponds to the minimum-weight
  vertex cover problem (see definition in \cite{garey1979cai}).
\end{proofsketch}

The minimum-weight vertex cover is NP-hard problem in general \cite{skiena}. However, for our specific case the sub-graph is a bi-partite graph in that no edges exist amongst the set of query nodes and the set of update nodes, but only between query and update nodes. Thus we can still solve the minimum-weight vertex cover problem by reducing it to the \emph{maximum network flow} problem \cite{Hochba97}. 
A polynomial time algorithm for the maximum network flow problem is the Edmonds-Karp algorithm \cite{corman1990ia}. This algorithm is based on the fact that a flow is maximum if and only if there is no augmenting path.  The algorithm
repeatedly finds an augmenting path and augments it with more flow,
until no augmenting path exists.

A primary challenges in employing an algorithm for the maximum network flow problem to \river is in (a) knowing the sequence of events in advance, and (b) extending it a limited size cache. The computation of the max-cover (or equivalently determining the max flow) changes as different knowledge of the future becomes available. In the example, if we just did not know what would happen at time $8s$, but knew everything
before, we would not load $o_4$, and instead ship $q_3$ and $q_7$.
Then when $q_8$ arrives, we would ship it too. Thus, different partial knowledge of
the future can lead to very different decisions and costs. The decisions obtained by the computation of the max-cover applies only to the objects in cache. We still need a mechanism for profitably loading objects in the cache.

\section{VCover}
\label{sec:vcover}

\vcover is an online algorithm for the data decoupling problem. It determines objects for which queries must be shipped and object for which updates must be shipped. The algorithm makes decisions in an ``online'' fashion, using minimal information about the incoming workload. In \vcover we also address the second challenge of a space-constrained cache. The overall algorithm is shown in Figure \ref{fg:servicequery}. 

The algorithm relies on two internal modules \updatemanager and \loadmanager to make two decisions: When a query arrives, \vcover first determines if the all objects accessed by the query are in cache. If all objects are in cache, 
this query is presented to the \updatemanager which chooses between shipping the outstanding updates required by the query, or shipping the query itself.  If, instead, a query arrives that accesses at least one object not in cache, then \vcover ships the query to the server, and also presents it to the \loadmanager which decides, in background, whether to load the missing object(s) or not.
We now describe the algorithms behind the \updatemanager and the \loadmanager.

\SetInd{0.5em}{0.5em}

\begin{algorithm}
\dontprintsemicolon
\linesnumbered
\SetVline
 {\bf \vcover}\;
{\bf Invocation:} By arriving query $q$, accessing objects $B(q)$,
with network traffic cost $\nu(q)$.\;
 {\bf Objective:} To choose between invoking \updatemanager, or shipping $q$ and invoking \loadmanager.\;
\BlankLine\;
\linesnumbered
\If{all objects in $B(q)$ are in cache}{
  \updatemanager with query $q$\;
}
\Else{
  Ship $q$ to server. Forward result to client\;
  In background: \loadmanager with query $q$\;
}
\caption{The main function in {\sf VCover}.}\label{fg:servicequery}
\end{algorithm}

\subsubsection{\updatemanager} \label{sec:updatemanager}

The \updatemanager builds upon the offline framework presented in Section \ref{sec:offline} to make ``online'' decisions as queries and updates arrive. The \updatemanager incorporates the new query into an existing internal interaction graph $G'$ by adding nodes for it and the updates it interacts  with and establishes the corresponding edges. To compute the minimum-weight vertex cover it uses the the Edmonds-Karp algorithm, which finds the maximum network flow in a graph. However, instead of running a network flow
 algorithm each time a query is serviced, it uses an incremental
 algorithm that finds just the change in flow since the previous
 computation (Figure \ref{fg:incre}).  If the result of the vertex cover computation is that
 the query should be shipped, then the \updatemanager does accordingly
 (Lines \ref{lnb:qinvc}--\ref{lne:qinvc}).  If not, then some updates
 are shipped.

\begin{algorithm}
\dontprintsemicolon
\linesnotnumbered
\SetVline
 {\bf \updatemanager on cache}\;
{\bf Invocation:} By \vcover with query $q$
accessing objects $B(q)$, with network traffic cost $\nu(q)$.\;
{\bf Objective:} To choose between shipping $q$ or
shipping all its outstanding interacting updates, and to update the
interaction graph.\;
\BlankLine\;
\linesnumbered
\If{each update interacting with $q$ has been shipped}{ \nllabel{lnb:incache}
  Execute $q$ in cache.  Send result to client\; \nllabel{lne:incache}
}
Add query vertex $q$ with weight $(\nu(q))$ to existing
internal interaction
graph $G'$ to get $G$\;  \nllabel{lnb:ig}
\ForEach{object $o \in B(q)$ marked \stale}{
  \ForEach{outstanding update $u$ for $o$}{
    Add update vertex $u$ to $G$, with weight its shipping cost, if not
    already present\;
  }
}
\ForEach{update vertex $u$ interacting with $q$}{
    Add edge $(u,q)$ to $G$\; \nllabel{lnse:ig}
}
Compute minimum-weight vertex cover $VC$ of $G$ using incremental network-flow
algorithm\; \nllabel{lnb:computevc}
\If{$q \in VC$}{ \nllabel{lnb:qinvc}
  \If{$q$ not already executed on cache}{
    Ship $q$ to server. Forward result to client\; \nllabel{lne:qinvc}
  }
}

\caption{{\sf UpdateManager} on cache.}\label{fg:updatemanager}
\end{algorithm}

\begin{figure}
\begin{enumerate}
  \item Let $G'$ be the previous internal interaction graph and $H'$
    the corresponding network constructed in previous iteration.
  \item Let $G$ be the current internal interaction graph.
  \item Add source and sink vertices to $G$ and corresponding
    capacitated edges as described in \cite{Hochba97} to construct
    network $H$.
  \item Find maximum flow in $H$, based on the known maximum flow for
    $H'$.
  \item Determine minimum-weight vertex for $G$ from maximum flow
    values of $H$, again as described in \cite{Hochba97}.
\end{enumerate}
\label{fg:incre}
\caption{Single iteration of the incremental network flow algorithm.}
\end{figure}

The key observation in computing the incremental max-flow is that as
vertices and edges are added and deleted from the graph there is no
change in the previous flow computation. Thus previous flow remains a
valid flow though it may not be maximum any more. The algorithm
therefore begins with a previous flow and searches for augmenting
paths that would lead to the maximum flow.  As a result, for any
sequence of queries and updates, the total time spent in flow
computation is not more than that for a single flow computation for the network corresponding to the entire
sequence.  If this network has $n$ nodes and $m$
edges, this time is $O(nm^2)$ \cite{corman1990ia}---much less
than the $O(n^2m^2)$-time the non-incremental version takes.

The \updatemanager does not record the entire sequence of queries and updates to make decisions. The algorithm always works on a \emph{remainder} subgraph
instead of the subgraph made over all the queries and updates seen so far. This remainder subgraph is formed from an existing
subgraph by excluding all update nodes picked in a vertex cover at any point,
and all query nodes not picked in the vertex cover. The exclusion can be safely done because
in selecting the cover at any point the
shipping of an update is justified based only on the past queries it
interacts with, and not with any future queries or updates and therefore will never be part of future cover selection.
This makes computing the cover robust to changes in workload. This technique also drastically reduces the size of the working subgraph making the algorithm very efficient in practice.

\subsubsection{Managing Loads}
\label{sec:loadmanager}
\vcover invokes the \loadmanager to determine if it is useful to load objects in the cache and save on
network costs or ship the corresponding queries. The decision is only made for those incoming queries which access
one or more objects not in cache. The difficulty in making this decision is due to queries accessing multiple objects,
each of which contributes a varying fraction to the query's total cost, $\nu(q)$. The objective is still to find, in an online fashion, the right combination of objects that should reside in cache such that the total network costs are minimized for these queries.

The algorithm in \loadmanager builds upon the popular Greedy-Dual-Size (GDS) algorithm \cite{CaoIra97} for Web-style proxy caching. GDS is an object caching algorithm that calculates an object's usage in the cache based on its frequency of usage, the cost of downloading it and its size. If the object is not being heavily used, GDS evicts the object. In GDS an object is loaded as soon it is requested. Such a loading policy can cause too much network traffic. In \cite{malik-bypss_cchng05}, it is shown that to minimize network traffic, it is important to ship queries on an object to the server until shipping costs equal to the object load costs have
been incurred.  After that any request to the object must load the object into the cache. The object's usage in the cache 
is measured from frequency and recency of use and thus eviction decisions can be made similar to GDS, based on object's usage in cache.

In \vcover queries access multiple objects and
the \loadmanager uses a simple twist on the algorithm in \cite{malik-bypss_cchng05} that still ensures
finding the right combination of choices. The \loadmanager (a)
randomly assigns query shipping costs among objects that a query
accesses, and (b) defers to GDS to calculate object's usage in the cache.
The random
assignment ensures that in expectation, objects are made
\emph{candidates} for load only when cost attributed equals the load
cost. A candidate for load is considered by considering in random sequence the set of objects accessed by the query (Line \ref{lnb:access} in Figure~\ref{fg:loadmanager}).
At this point, the shipping cost could be attributed to the object.

The load manager employs two techniques that makes its use of Greedy-Dual-Size more efficient. We explain the need for the techniques:
\begin{itemize}
\item Explicit tracking of the total cost (due to all queries) of each object is inefficient.
The load manager eliminates explicit tracking of the total cost
by using randomized loading (Lines \ref{lnb:ra}-\ref{lne:ra}). 
This leads to a more space-efficient implementation of the algorithm
in which counters on each object are not maintained. When the
shipping cost from a single query covers the entire load cost of an
object, the object is immediately made a candidate to be loaded.
Else, it is made so with probability equal to the ratio of the cost
attributed from the query and the object's load cost. 
\item  In a given subsequence
of load requests generated by $q$, say $o_1,o_2,\ldots,o_m$, it is
possible that the given $\aobj$ loads an $o_i$ only to evict it to
accommodate some later $o_j$ in the \emph{same} subsequence.  Clearly,
loading $o_i$ is not useful for the \loadmanager. To iron out such inefficiencies we use a lazy
version of $\aobj$, in our case  Greedy-Dual-Size. 
\end{itemize}

In the \loadmanager once the object is loaded, the system ensures that all updates that came while the object was being loaded are applied and the object is marked fresh by both cache and server.

\SetInd{0.5em}{0.5em}
\begin{algorithm}
\dontprintsemicolon
\linesnumbered
\SetVline
 {\bf \loadmanager on cache}\;
{\bf Invocation:} By \vcover with query $q$, accessing objects $B(q)$,
with network traffic cost $\nu(q)$.\;
 {\bf Objective:} To load useful objects\;

\BlankLine\;
\linesnumbered
$c\leftarrow \nu(q) $\; \nllabel{lnb:ra}
\While{$B(q)$ has objects not in cache and $c>0$}{
  $o \leftarrow$ some object in $B(q)$ not in cache\; \nllabel{lnb:access}
  \If(\tcc*[f]{$l(o)$ is load cost of $o$}){$c \geq l(o)$}{
    $\Aobj\_\lazy$ with input $o$\; \nllabel{lnb:a1}
    $c \leftarrow c-l(o)$\;
  }
  \Else(\tcc*[f]{randomized loading}) {
    With probability $c/l(o)$: $\Aobj\_\lazy$ with input $o$\; \nllabel{lnb:a2}
    $c\leftarrow 0$\;  \nllabel{lne:ra}
  }
  Load and evict objects according to $\Aobj\_\lazy$\; \nllabel{lnb:a3}
  \If{$\Aobj\_\lazy$ loads $o$}{
    Both server and cache mark $o$ \fresh\;
  }
}
\caption{{\sf LoadManager} on cache given an object caching algorithm
  \Aobj. }\label{fg:loadmanager}
\end{algorithm}

\vspace{5pt}
\noindent{\bf Discussion:} There are several aspects of \vcover that we would like to highlight.
First, in \river we have focused on reducing network traffic. Consequently, in presenting \vcover 
we have included only those decisions that reduce network traffic. 
These decisions naturally decrease response times of queries that access objects in cache. 
But queries for which updates need to be applied may be delayed. 
In some applications, such as weather prediction, which have similar rapidly-growing repositories, 
minimizing overall response time is equally important. 
To improve the response time performance of delayed queries, some updates can be \emph{preshipped}, \emph{i.e.,} proactively sent by the server. For lack of space we have omitted how preshipping in \vcover can further improve overall response times of all queries. We direct the reader to the accompanying technical report \cite{malik09river} for this.

In the \river architecture data updates correspond predominantly to data inserts (Section \ref{sec:problemformulation}).
This is true of scientific repositories. However the decision making in \vcover is independent of an update specification 
and can imply any of the data modification statements \emph{viz.} insert, delete or modify. Thus we have chosen 
the use of the term update. 

\loadmanager adopts a randomized mechanism for loading objects. This is space-efficient as it 
obviates maintaining counters on each object. This efficiency is motivated by meta-data issues 
in large-scale remote data access middlewares that cache data from several sites \cite{malik09river}.

Finally, an implementation of \vcover requires a semantic framework that determines the mapping between the query, $q$, and the data objects, $B(q)$, it accesses. The complexity of such a framework depends upon the granularity at which data objects are defined. If the objects are tables or columns, such a mapping can be determined by the specification of the query itself. If the objects are tuples, finding such a mapping a-priori is difficult. However for most applications such a mapping can be found by exploiting the semantics. For instance, in astronomy, queries specify a spatial region and objects are also spatially partitioned. Thus finding a mapping can be done by some pre-processing as described in Section \ref{sec:exp}.

\section{Benefit: An alternative approach to the decoupling problem}
\label{sec:benefit}

In \vcover we have presented an algorithm that exploits the 
combinatorial structure (computing the cover) within the data decoupling problem 
and makes adaptive decisions using ideas borrowed from
online algorithms \cite{Borodin:1998:OCC}. An alternative approach to solving the data decoupling problem is 
an exponential smoothing-based algorithm that makes decisions based on heuristics.
We term such an algorithm \benefit as it is inherently greedy in its decision making.

In \benefit we divide the sequence of queries and updates into windows
of size $\delta$.  At the beginning of each new window $i$, for each
object currently in the cache, we compute the ``benefit'' $b_{i-1}$
accrued from keeping it in the cache during the past window $(i-1)$.
$b_{i-1}$ is defined as the network cost the object saves by answering
queries at the cache, less the amount of traffic it causes by having
updates shipped for it from the server.  Since a query answered at the
cache may access multiple objects, the cost of its shipping (which is
saved) is divided among the objects the query accesses in proportion to
their sizes.  This form of dividing the cost has been found useful in
other caches as well \cite{malik-bypss_cchng05,Bagchi:2006:ACE}.

For each object currently not in cache, we compute similarly the
benefit it would have accrued if it \emph{had been} in the cache
during window $(i-1)$.  But here, we further reduce the benefit by the
cost to load the object.

A forecast $\mu_i$ of the benefit an object will accrue during the next
window $i$ is then computed using exponential smoothing: $\mu_i =
(1-\alpha)\mu_{i-1} + \alpha b_{i-1}$, in which $\mu_{i-1}$ was the
forecast for the previous window, and $\alpha$, which is $0 \leq \alpha
\leq 1$, is a learning parameter.

We next consider only objects with positive $\mu_i$ values and rank
them in decreasing order.  For window $i$, we greedily load objects
in this order, until the cache is full.  Objects which were already
present in cache in window $(i-1)$ don't have to be reloaded.
For lack of space, pseudo-code of \benefit is not presented in this paper, 
but is included in the accompanying technical report.

\benefit reliance on heuristics is similar to previously proposed algorithms for online view materialization \cite{labrinidis-dt_frshnss_wb_srvrs04,LR00}.  It,
however, suffers from some weaknesses.  The foremost is that \benefit
ignores the combinatorial structure within the problem by dividing the
cost of shipping a query among the objects the query accesses in
proportion to their sizes. This difference is significant in that \benefit's performance can be proved analytically. However, in \vcover the combinatorial structure leads to a mathematical bound on its performance. 
For details on the proof of performance, the reader is directed to the technical report \cite{malik09river}. 
Further, \benefit's decision making is heavily dependent on the size of window chosen.
It also needs to maintain state for each object (the $\mu_i$ values, the queries which access it,
etc.) in the database irrespective of whether it is in the cache or
not.  

\section{Empirical Evaluation}
\label{sec:exp}
\label{sec:experiments}


In this section, we present an empirical evaluation of \scicache on real astronomy
query workloads and data. The experiments validate using a data decoupling 
framework for minimizing network traffic. 
Our experimental results show that \scicache (using \vcover) reduces the traffic by
nearly half even with a cache that is one-fifth the size of the
server.  
\vcover
outperforms \benefit by a factor that varies between 2-5 under
different conditions.  

\subsection{Experimental Setup}

{\bf Choice of Survey:} Data from rapidly growing repositories, such as those of the Pan-STARRS and the LSST, is currently unavailable for public use.
Thus we used the data and query traces from the SDSS for experimentation. SDSS periodically publishes updates via new data release. To build a rapidly growing SDSS repository, we simulated an update trace for SDSS in consultation with astronomers. The use of SDSS data in validating \river is a reasonable choice as the Pan-STARRS and the LSST databases have a star-schema similar to the SDSS. When open to public use, it is estimated that 
the Pan-STARRS and the LSST repositories will be queried similar to the SDSS. 

{\bf Setup:} Our prototype system consists of a server database and a middleware cache database, both implemented on an 
IBM workstation with 1.3GHz Pentium III processor and 1GB of memory, running Microsoft Windows, each running 
a MS SQL Server 2000 database. A sequence of data updates are applied to a server database 
Queries, each with a currency specification criteria, arrive concurrently 
at a cache database. To satisfy queries with the latest data, the server and cache database uses MS SQL Server's 
replica management system to ship queries and updates, and, when necessary, it uses bulk copying for object loading. 
The decisions of when to use 
each of the data communication mechanism is dictated by the optimization framework of \river, 
implemented as stored procedures on the server and cache databases.

{\bf Server Data:} The server is an SDSS database partitioned in spatial data objects. To build spatial data objects, we use the primary table in SDSS, called the PhotoObj table, which stores data about each astronomical body including its spatial location and about 700 other physical attributes. The size of the table is roughly 1TB. The table is partitioned using a recursively-defined quad tree-like index,
called the \emph{hierarchical triangular mesh} \cite{szalay-hrrchcl_trnglr_msh}. The \texttt{HTM} index 
conceptually divides the sky into partitions; in the database these partitions translate into roughly
equi-area data objects. A partitioning of the sky and therefore the data depends upon the level of the 
HTM index chosen. 
For most experiments we used
a level that consisted of 68 partitions (ignoring some which weren't
queried at all), containing about 800 GB of data.  These partitions
were given object-IDs from 1 to 68. Our choice of 68 objects is based on a cache granularity experiment explained in 
Section \ref{sec:results}. The The data in each object varies from
as low as 50 MB to as high as 90 GB. 

{\bf Growth in Data:} Updates are in the form of new data inserts and are applied to a spatially defined data object. 
We simulated the expected update patterns of the newer astronomy surveys under consultation with
astronomers. Telescopes collect data by scanning specific regions of
the sky, along great circles, in a coordinated and systematic fashion
\cite{SGT+02}.  Updates are thus clustered by regions on the
sky.  Based on this pattern, we created a workload of 250,000 updates. The size of an update 
is proportional to the density of the data object.

{\bf Queries:} We extracted a query workload of about 250,000 queries received from January to February, 2009. 
The workload trace consists of several kinds of queries, including
range queries, spatial self-join queries, simple selection queries, as
well as aggregation queries. Preprocessing on the traces involves removing
queries that query the logs themselves. In this trace, 98\% of the queries and 99\% of the network traffic 
due to the queries is due to the PhotoObj table or views defined on PhotoObj. 
Any query which does not query the PhotoObj is bypassed and shipped to the server.

{\bf Costs: } The traffic cost of shipping a query is the actual number of bytes in
its results on the current SDSS database.  The traffic cost of
shipping an update was chosen to match the expected 100 GB of
update traffic each day in the newer databases.


\begin{figure}[h]
\centering
\includegraphics[width=0.49\textwidth]{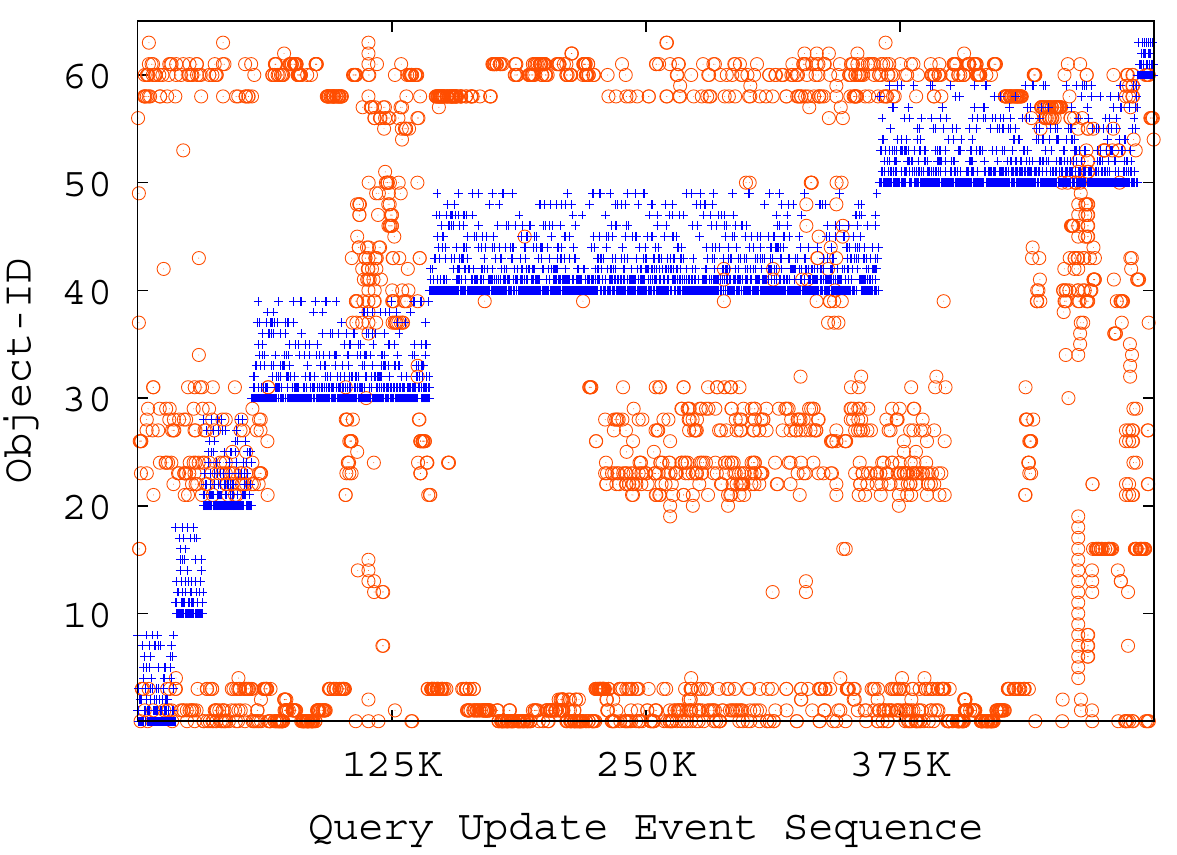}
\includegraphics[width=0.49\textwidth]{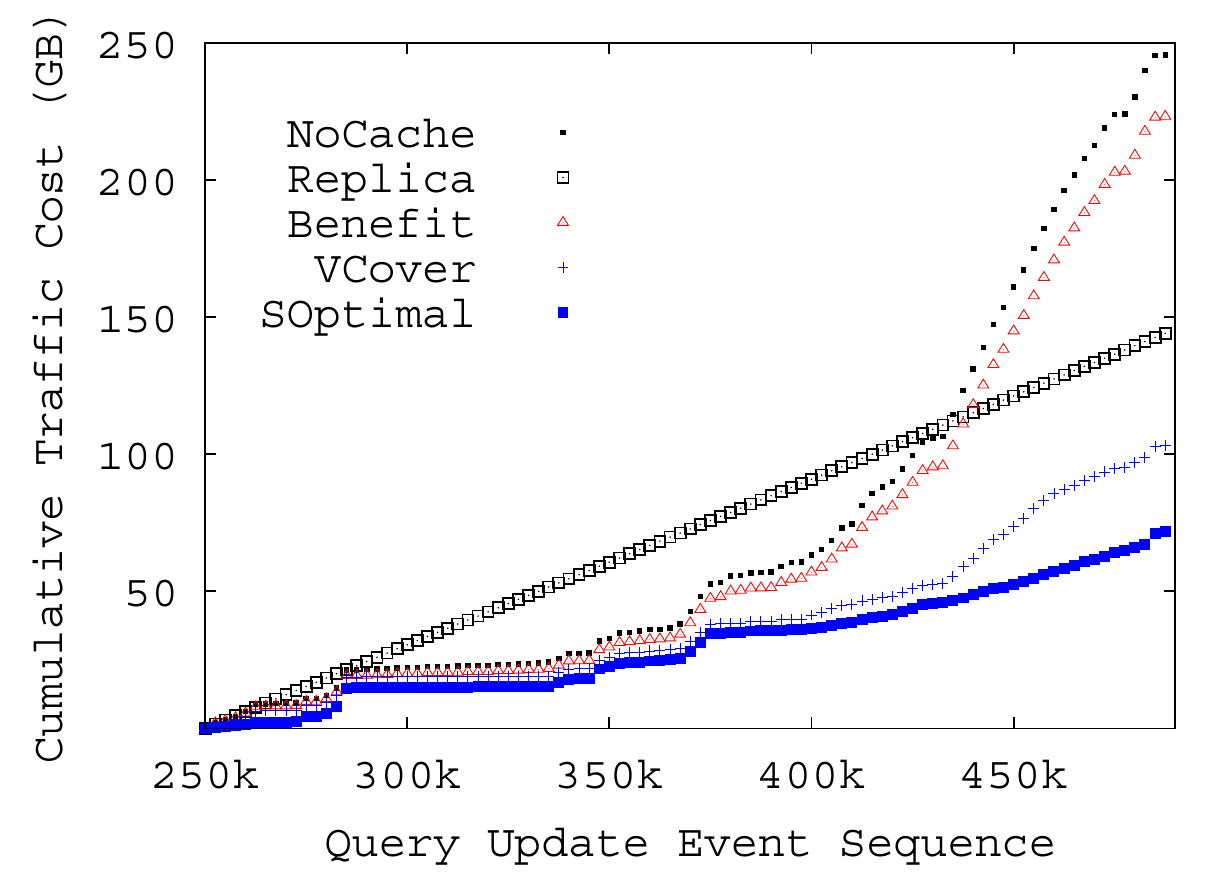}
\caption{(a) Object-IDs corresponding to each query (red ring) or update (blue cross)
  event. Queries evolve and cluster around different objects over time.
(b) Cumulative traffic cost.  \vcover, almost as good as \soptimal,
  outperforms others. } 
  \label{fg:ac}
\label{fg:sdsscost}
\end{figure}

A sample of the query and update event sequence is shown in Figure
\ref{fg:ac}.  The sequence is along the $x$-axis, and for each event,
if it is an update, we put a blue diamond next to the object-ID affected
by the update.  If the event is a query, we put a yellow dot next to
\emph{all} object-IDs accessed by the query.  Even though this rough
figure only shows a sample of the updates and queries, and does not
include the corresponding costs of shipping, it supports what we
discovered in our experiments: object-IDs 22, 23, 24, 62, 63, 64 are
some of the query hotspots, while object-IDs 11, 12, 13, 30, 31, 32
are some of the update hotspots.  The figure also indicates that
real-world queries do not follow any clear patterns.


\subsubsection*{Two Algorithms, Three Yardsticks}

We compare the performances of \vcover, the core algorithm in \river with 
\benefit, the heuristic algorithm that forecasts using exponential smoothing.  
We also compare the performance of these two algorithms with three other policies. These policies act as yardsticks
in that the algorithm can be considered poor or excellent if it performs better or worse than these policies. The policies are:
\squishlist
\item {\sf \bf NoCache}: Do not use a cache.  Ship all queries to the
  server.  Any algorithm (in our case \vcover and \benefit), which has a performance worse than \nocache is
  clearly of no use.
\item {\sf \bf Replica}: Let the cache be as large as the server and
  contain all the server data. To satisfy queries with the most current data, 
  ship all updates to the cache as soon as they arrive at server. If \vcover and \benefit, both of which respect a 
  cache size limitation, perform better than \replica they are clearly good.
\item {\sf \bf SOptimal}: Decide on the best \emph{static set of
    objects} to cache \emph{after} seeing the entire query and update sequence.  
 Conceptually, its decision is equivalent to the single decision of
 \benefit using a window-size as large as the entire sequence, but in an offline manner.
  To implement the algorithm we 
  loads all objects it needs at the beginning and do not ever evict
  any object. As updates arrive for objects in cache they are shipped.
  Any online algorithm, which cannot see the queries and updates in advance, but with performance close to the \soptimal is
  outstanding.
\squishend

\aparagraph{Default parameter values, warmup period} Unless specified
otherwise, in the following 
experiments we set the cache
size to 30\% of server size,
and the window size $\delta$ in \benefit to 1000.
The choice are obtained by varying the parameters in the experiment to obtain the optimal value.   
The cache undergoes an initial warm-up period of about 250,000 events for both
\vcover and \benefit. A large warm-up period is a characteristic of this particular 
workload trace in which queries with small query cost occur earlier in trace. 
As a result objects have very low probability of load. 
In this warm-up period the cache remains nearly empty and
almost all queries are shipped. In general, our experience with other workload traces of similar size 
have shown that a range of a warm period can be anywhere from 150,000 events to 300,000 events. 
To focus on the more interesting post warm-up period we do not show the events and the costs incurred during
the warmup period.

\subsection{Results}
\label{sec:results}

\aparagraph{Minimizing traffic cost} The cumulative network traffic cost along the query-update event sequence
for the two algorithms and three yardsticks is in Figure \ref{fg:sdsscost}(a). \vcover is
clearly superior to \nocache and \benefit: as more query and update events arrive it continues to perform better than them and at the end of the trace has an improvement by a factor of at least 2 with \benefit, 
and to \replica by a factor about 1.5. (For replica load costs and
cache size constraints are ignored.) Further, as more data intensive queries arrive, \benefit is barely better than \nocache.  \vcover closely follows \soptimal until about Event 430K, when it diverges, leading to a final cost about 40\% (35 GB)
higher.  On closer examination of the choices made by the algorithm,
we discovered that, with hindsight, \soptimal determines that
Object-ID 39 of 38 GB and Object-ID 29 of 4 GB would be useful to
cache and loads them at the beginning.  But \vcover discovers it only
after some high-shipping cost queries that arrive for these objects,
and loads them around Event 430K, thus paying both the query
shipping cost and load cost. 

\begin{figure}[h]
\centering
\includegraphics[width=0.49\textwidth]{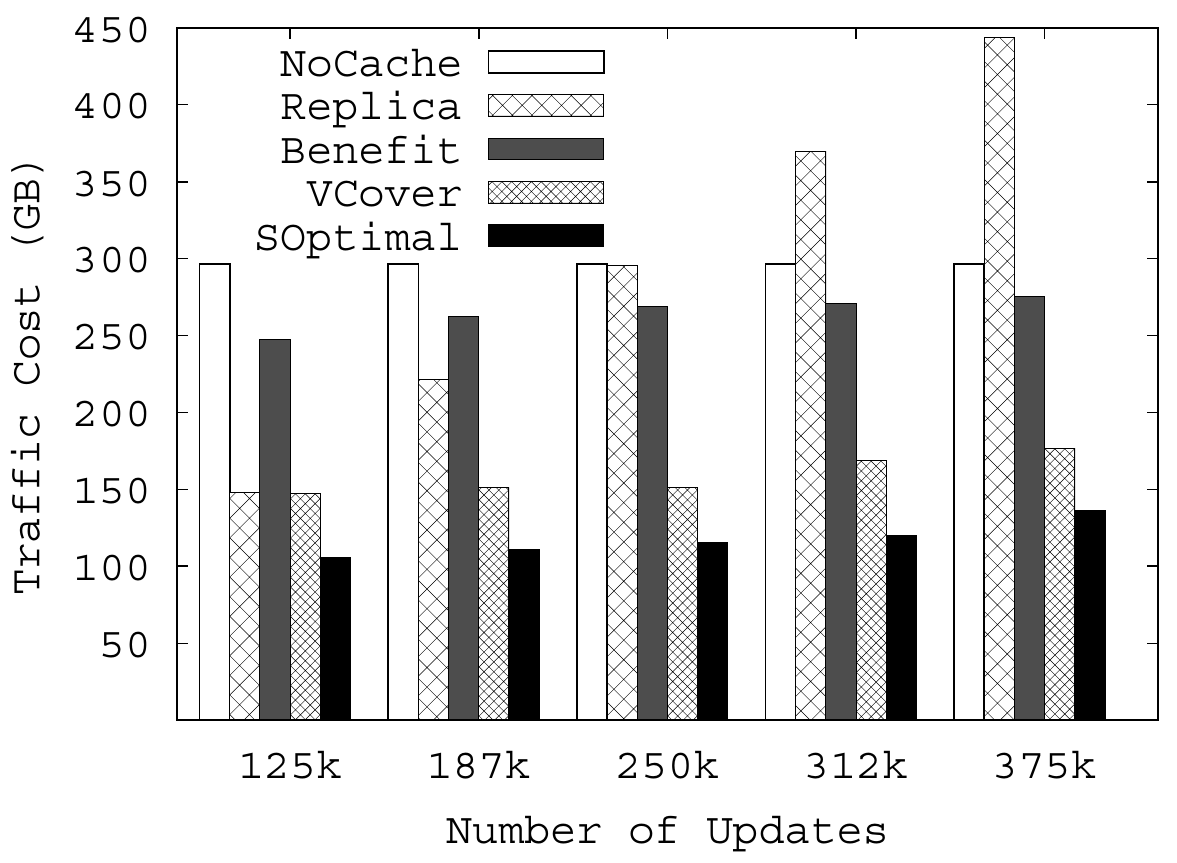}
\includegraphics[width=0.49\textwidth]{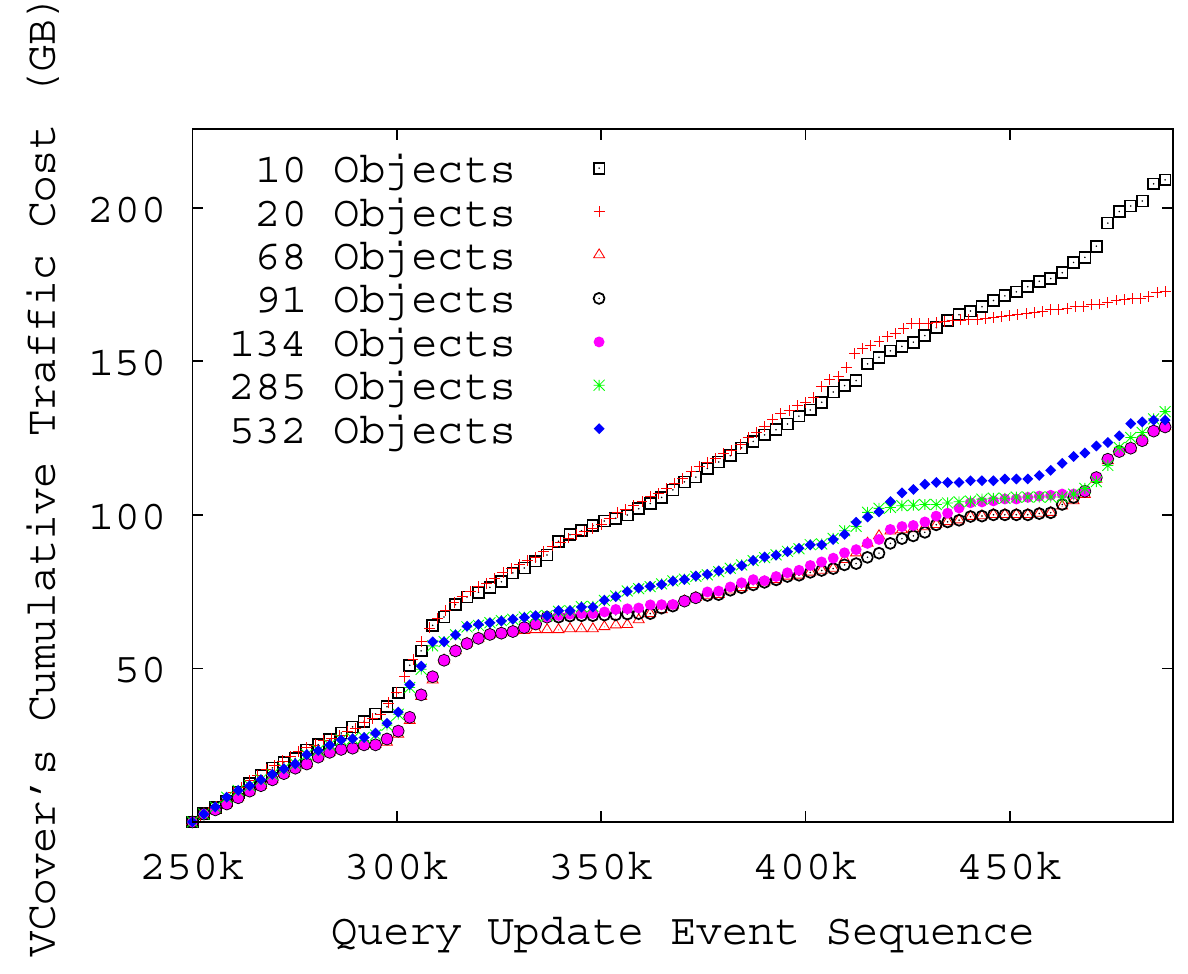}
\caption{(a) Traffic cost for varying number of updates. (b) \vcover's cumulative traffic cost for different choices of
  object sets.}
\label{fg:varyingnoofupdates}
\label{fg:gran}
\end{figure}

\aparagraph{Varying number of updates} A general-purpose caching
algorithm for dynamic data should maintain its performance in the face
of different rates of updates and queries.  In Figure
\ref{fg:varyingnoofupdates}(b) we plot the final traffic cost for each
algorithm for different workloads, each workload with the same 250,000
queries but with a different number of updates.  The simplistic
yardstick algorithms \nocache and \replica do not take into account
the relative number of updates and queries.  Since the queries remain
the same, the cost of \nocache is steady at 300 GB.  But as the number
of updates increase the cost of \replica goes up: the three-fold
increase in number of updates results in a three-fold increase in
\replica's cost.  The other three algorithms, in contrast, show only
a slight increase in their cost as the number of updates increase.
They work by choosing the appropriate objects to cache for each
workload, and when the updates are more, they compensate by keeping
fewer objects in the cache.  The slight increase in cost is due to
query shipping cost paid for objects which are no longer viable to
cache.  This experiment illustrates well the benefits of \scicache
irrespective of the relative number of queries and updates.

\aparagraph{Choice of objects} In Figure \ref{fg:gran}(a) we plot
\vcover's cumulative traffic cost, along the event sequence, for
different choices of data objects.  Each choice corresponds to a different
level in the quad tree structure in SDSS, from a set of 10 objects
corresponding to large-area upper-level regions, to a set of 532
objects corresponding to small-area lower-level regions.  Each set
covers the entire sky and contains the entire data.  The performance
of \vcover improves dramatically as number of number of objects
increase (sizes reduce) until it reaches 91 and then begins to
slightly worsen again.  The initial improvement is because as objects
become smaller in size, less space in the cache is wasted, the hotspot
decoupling is at a finer grain, and thus more effective.  But this
trend reverses as the objects become too small since the likelihood
that future queries are entirely contained in the objects in cache
reduces, as explained next.  It has been observed in scientific
databases that it is more likely that future queries access data which
is ``close to'' or ``related to,'' rather than the exact same as, the
data accessed by current queries \cite{malik-bypss_cchng05}.  In
astronomy, e.g., this is partly because of several tasks that scan the
entire sky through consecutive queries.

\section{Conclusion}
\label{sec:conclusion}

Repositories in data-intensive science are growing rapidly in size. 
Scientists are also increasingly 
interested in the time dimension and thus demand latest data to be part of query results.  
Current caches provide minimal support for incorporating the latest data at the repository; many of them 
assume repositories are static. This often results in runaway network costs. 

In this paper we presented \river\ a dynamic data middleware cache system for
rapidly growing repositories.  \river is based on a data decoupling framework---it separates objects that are rapidly growing from objects that are heavily queried. By effective decoupling the framework naturally minimizes network costs. 
\river relies on \vcover, a robust, adaptive algorithm that decouples data objects by examining the cost of usage and currency requirements. \vcover's decoupling is based on sound graph theoretical principles making the solution nearly optimal over evolving scientific workloads. We compare the performance of \vcover with \benefit, a greedy heuristic, which is commonly employed in dynamic data caches for commercial applications. Experiments show that \benefit scales poorly than \vcover on scientific workloads in all respects.

A real-world deployment of \river would need to also
consider several other issues such as reliability, failure-recovery,
and communication protocols.  
The future of applications, such as the Pan-STARRS and the LSST depends on scalable
network-aware solutions that facilitate access to data to a large number of
users in the presence of overloaded networks.  \river is a step
towards meeting that challenge.

\aparagraph{\bf Acknowledgments} The authors sincerely thank Alex Szalay for 
motivating this problem to us, and Jim Heasley for describing 
network management issues in designing database repositories for the
Pan-STARRS survey. Ani Thakar acknowledges support from the Pan-STARRS project.

{\small
\bibliographystyle{abbrv}

\bibliography{dache}
}

\end{document}